\documentclass[12pt,preprint]{aastex}

\newcounter{saveeqn}
\newcommand{\alpheqn}{\setcounter{saveeqn}{\value{equation}}%
  \stepcounter{saveeqn}\setcounter{equation}{0}%
  \renewcommand{\theequation}
      {\mbox{\arabic{saveeqn}\alph{equation}}}}
\newcommand{\reseteqn}{\setcounter{equation}{\value{saveeqn}}%
  \renewcommand{\theequation}{\arabic{equation}}}

\shorttitle{Rotating Galaxy Clusters in SDSS and 2dFGRS}
\shortauthors{Hwang \& Lee}

\begin{document}

\title{Searching for rotating galaxy clusters in SDSS and 2dFGRS}

\author{Ho Seong Hwang and Myung Gyoon Lee}
\affil{Astronomy Program, Department of Physics and Astronomy, Frontier
Physics Research Division,
Seoul National University, 56-1 Sillim 9-dong, Gwanak-gu, Seoul 151-742, Korea}
\email{hshwang@astro.snu.ac.kr, mglee@astrog.snu.ac.kr}

\begin{abstract}
We present a result of searching for galaxy clusters
that show an indication of global rotation using a spectroscopic sample of galaxies
in SDSS and 2dFGRS. 
We have determined the member galaxies of 899 Abell clusters covered in SDSS and 2dFGRS
using the redshift and the positional data of galaxies,
and have estimated the ratio of the cluster rotation amplitude to the cluster velocity dispersion
and the velocity gradient across the cluster. 
We have found 12 tentative rotating clusters 
that have large ratios of rotation amplitude to dispersion and large velocity gradients.
We have determined the morphological parameters for 12 tentative rotating clusters using the positional
information of the member galaxies: the ellipticity of the dispersion ellipse 
is in the range of 0.08$-$0.57, and the position angle of major or minor axis does not appear
to be related to the position angle of rotation axis.
We have investigated the substructures in the sample of tentative rotating clusters, 
finding from the Dressler-Shectman plots that the majority (9 out of 12) of
clusters show an evidence of substructure due to the spatially correlated
velocities of galaxies.
We have selected six probable rotating clusters 
(Abell 0954, Abell 1139, Abell 1399, Abell 2162, Abell 2169, and Abell 2366)
that show a single number density peak around
the cluster center with a spatial segregation of the high and low velocity galaxies. 
We have found no strong evidences of a recent merging for the probable rotating clusters:
the probable rotating clusters do not deviate significantly from the relation of 
the X-ray luminosity and the velocity dispersion or the virial mass of the clusters,
and two probable rotating clusters (A0954 and A1399) have small values of the peculiar velocities and 
the clustercentric distances of the brightest cluster galaxies.

\end{abstract}



\section{Introduction}


Global rotation of galaxy clusters has been suggested for several galaxy clusters 
(e.g. \citealt{mh83,oh92,db01a}),
but there are still no conclusive evidences of cluster rotation. 
There are two kinds of objects to consider with regard to the rotation of galaxy
clusters: intracluster medium (ICM) and the galaxies.

For ICM, recent studies have searched for the bulk motion or the velocity gradient of ICM 
in several clusters using {\it ASCA} and {\it Chandra} data.
\citet{db01a,db01b} directly detected, for the first time, ICM bulk motions in the
outer regions of the Centaurus (Abell 3526) and the Perseus (Abell 426) clusters using {\it ASCA} data.
Later, \citet{db05} identified two more clusters (Abell 576 and RX J0419.5+0225) that have
the most significant velocity gradients among 12 clusters for which
velocity mapping could be performed with useful precision in the {\it ASCA} archive.
Recently, \citet{db06} measured the velocity difference of (2.4$\pm$1.0)$\times 10^{3}$ km s$^{-1}$ 
for the regions opposed around the center of ICM in the Centaurus cluster using {\it Chandra} data. 
If this velocity difference is due to the gas circulation around the cluster center, then the corresponding circular velocity
will be as large as (1.2$\pm$0.7)$\times 10^{3}$ km s$^{-1}$.

For the galaxies in clusters, the global velocity gradient suggestive of cluster rotation
has been detected in some clusters \citep{mh83,biviano96,dHK96,burgett04}.
\citet{dHK96} found that 13 clusters show significant velocity gradients 
in the range of 240$-$1230 km s$^{-1}$ Mpc$^{-1}$ among 72 clusters.
\citet{burgett04} reported that three clusters (Abell 1139, Abell 1663, and Abell S333)
exhibit velocity-position characteristics consistent with the presence of possible cluster
rotation, shear, or infall dynamics.
Up to now, there is only one galaxy cluster of which global rotation of galaxies was studied in detail: Abell 2107.
\citet{oh92} found from the analysis of 75 member galaxies that 
Abell 2107 appears to be a regular cluster (single peaked galaxy distribution,
regular X-ray morphology, and Gaussian velocity histogram), but has spatially correlated
velocities of galaxies (high and low velocity galaxies are segregated spatially).
They concluded that
the spatial distribution of galaxy velocity is consistent with a rotation of a single cluster at
98 \% confidence level.
Recently, \citet{kalinkov05} suggested that the virial mass of Abell 2107 determined 
in their study, should be reduced
from (3.2$\pm$0.6)$\times 10^{14} M_\odot$ to (2.8$\pm$0.5)$\times 10^{14} M_\odot$ 
due to the correction for the cluster rotation. 

Theoretically, a galaxy cluster is expected to acquire angular momentum
through an off-axis merging or a global rotation of the universe.
First, off-axis merging between two clusters is expected 
to provide angular momentum to the clusters, 
resulting in rotating clusters \citep{ricker98,takizawa00,rs01,pawl05}.
According to off-axis merging models,
the global rotation of ``collisional'' ICM originated by merging
survives longer than that of ``collisionless'' galaxies in
clusters \citep{rf00}. Therefore, 
it is expected that it is more difficult to detect the global rotation
of galaxies than that of ICM.
Secondly, the global rotation of the universe may provide angular momentum
to the celestial bodies on their formation, resulting in rotating systems \citep{li98}.
Li model explains the empirical relation between the angular momenta and the masses of galaxies. 
Later, \citet{god03} showed that Li model predicts the presence of a minimum (i.e. vanishing angular momentum)
in the relation between the angular momenta and the masses of celestial bodies.
Their prediction was observationally tested by \citet{god05}; vanishing angular momenta
for the masses corresponding to the galaxy groups and non-vanshing angular momenta for smaller
and larger structures (compact groups and rich galaxy clusters). However, the non-vanishing angular momenta
for galaxy clusters were found indirectly based on the study of the galaxy orientation. Therefore, it is needed to
probe the angular momenta of galaxy clusters directly based on dynamical analysis.

Recently, galaxy redshift surveys such as the Sloan Digital
Sky Survey (SDSS, \citealt{york00}) and the 2dF Galaxy Redshift Survey
(2dFGRS, \citealt{col01}) have provided redshift data for large samples of
galaxies. 
Therefore, by identifying member galaxies in clusters using the redshift and the positional data,
we can search for the global rotation
of galaxies in clusters for a large sample of galaxy clusters in these data.


In this paper, we present a result of searching for galaxy clusters
that show an indication of global rotation using a spectroscopic sample of galaxies
in SDSS and 2dFGRS. Section 2 describes
the galaxy sample and the cluster sample. The algorithm
for identifying tentative rotating clusters and the results are presented in Section 3.
Analysis of the cluster morphology and of the substructure for tentative rotating clusters
is given in Section 4. A detailed analysis of the individual clusters is presented in Section 5.
We discuss the properties of the probable rotating clusters in Section 6.
A summary of this study is given in the final section. Throughout this paper, we adopt
cosmological parameters $h=0.7$, $\Omega_{\Lambda}=0.7$, and
$\Omega_{M}=0.3$.

\section{Data}\label{data}
We used the data for the spectroscopic sample of galaxies 
in the Fifth Data Release of SDSS\footnote{The details
of Data Release 5 can be found on the SDSS web site (http://www.sdss.org/dr5/).}, 
and in the Final Data Release of 2dFGRS \citep{col01}. 

The SDSS Data Release 5 contains five-band ({\it u, g, r, i, z}) photometric
data for 215 million objects over 8,000 deg$^2$, and optical spectroscopic
data for one million objects of galaxies, quasars, and stars over 5,740 deg$^2$. 
We used the spectroscopic sample of 674,000 galaxies for this study. 
The median redshift for the spectroscopic sample of galaxies is 0.11.
The redshifts were measured from the spectra covering 3800$-$9200 $\AA$ 
with an uncertainty of $\sim$30 km s$^{-1}$. The redshift
confidence parameter (zConf) was assigned from 0 to 1 in the SDSS catalogs. 
We used only the galaxies with zConf $\ge$ 0.65.

The 2dFGRS contains the spectra for 246,000 galaxies
selected in the photographic $b_{\rm J}$ band from the APM galaxy
catalog. The survey covers over 2,000 deg$^2$ and, the median redshift for the sample of galaxies
is similar to that of SDSS. The redshift measurements were done from
the spectra covering 3600$-$8000 $\AA$, and the redshift quality
parameter Q was assigned in the range of 1$-$5. The redshift measurements with Q$\ge$3 are 98.4 \%
reliable, and have an overall rms uncertainty of $\sim$85 km
s$^{-1}$. We used only the galaxies with Q $\ge$ 3
in this study.

Since some galaxies were covered in both SDSS and 2dFGRS, we matched
the galaxies found in SDSS with those in 2dFGRS to make a master catalog of galaxies.
The mean difference of radial velocities between the measurements of SDSS and 2dFGRS 
for 31,300 matched galaxies is $\Delta v (=v_{\rm SDSS}-v_{\rm 2dFGRS})\sim$13 km s$^{-1}$. 
We corrected the velocities of galaxies in SDSS by the mean difference of radial velocities,
and used the average value of the velocities measured in SDSS and in 2dFGRS for further analysis. 

We used the Abell catalog of galaxy clusters \citep{aco89} to find the clusters in the survey data.
Among the Abell clusters, we selected the clusters that have known spectroscopic redshifts
in NASA/IPAC Extragalactic Database (NED). Finally, we selected 899 clusters 
located within
the survey regions of SDSS and 2dFGRS as a sample of clusters for further analysis.

\section{Identification of rotating cluster candidates}\label{rotating}
\begin{figure}
\includegraphics [width=165mm] {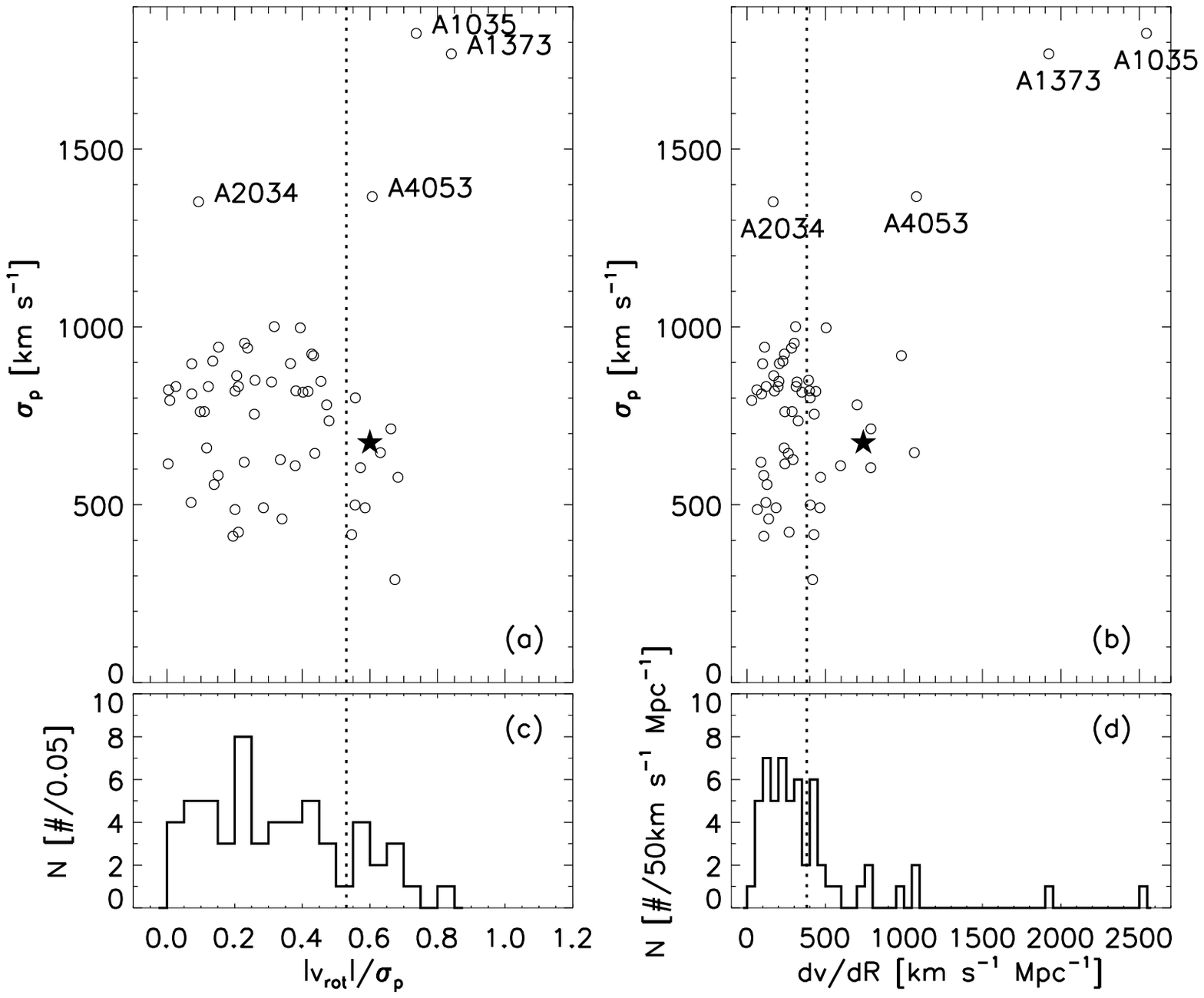}
\caption{Velocity dispersions of galaxy clusters
versus the ratios of the absolute value of rotation amplitude to the velocity dispersion (a),
and versus the global velocity gradients (b). The histograms for the ratios and for the velocity gradients
are shown in (c) and (d), respectively. The dotted vertical lines indicate the adopted selection criteria
for the tentative rotating cluster candidates. 
Star symbols represent A2107, known as a probable rotating cluster.} \label{fig-rotcand}
\end{figure}

We applied the following procedure to identify the
rotating cluster candidates in SDSS and 2dFGRS:

\begin{enumerate}
\item In order to select the member galaxies in the target clusters, we used
``shifting gapper'' method of \citet{fadda96}. 
In the plot of radial velocity versus clustercentric distance of galaxies for a given cluster,
we selected the galaxies as cluster members using a velocity gap of 950 km s$^{-1}$ 
and a distance bin of 0.2 Mpc shifting along the clustercentric distance from the cluster center. 
We used a larger bin width if the number of galaxies in a bin is less than 15.
We applied the method to the galaxies
within the radius at which the distance between adjacent galaxies is larger than 0.1 Mpc.
We iterated the procedure until the number of cluster members becomes stable.  
Finally, we selected 56 galaxy clusters in which the number of
member galaxies is greater than or equal to 40 for further analysis.

\item To investigate the global rotational property of galaxy clusters,
we fit the observed radial velocities $v_p$ of the cluster galaxies with
a function of a position angle,

\begin{equation}
v_p (\Theta) = v_{\rm sys} + v_{\rm rot} \sin(\Theta - \Theta_0)\ ,
\label{eq-rot}
\end{equation}

where $\Theta$ is the projected position angle of a galaxy relative to
the cluster center (measured from North to East), $\Theta_0$ is the
projected position angle of the rotation axis of the cluster, 
$v_{\rm rot}$ is the rotation amplitude, and
$v_{\rm sys}$ is the systemic velocity of the galaxy cluster.

Similarly, the effect of the global rotation of galaxy clusters can appear as a
velocity gradient across the cluster. To investigate the global velocity
gradient, we fit the observed radial velocities of the cluster galaxies with
a function of a position in the plane of the sky,

\begin{equation}
v_p (X,Y) = v_{\rm sys} + \frac{\partial v}{\partial X} X+ 
            \frac{\partial v}{\partial Y} Y,
\label{eq-fit}
\end{equation}

where X and Y are clustercentric distances in the direction of right ascension
and declination, respectively. 
We fit the observed radial velocities of cluster galaxies using equation (\ref{eq-rot}) and (\ref{eq-fit})
with $v_{\rm sys}$ as a fixed value of the mean radial velocity of cluster galaxies (the biweight
location of \citealt{beers90}).

In Figure \ref{fig-rotcand}, we plot the velocity dispersions (the biweight scale of \citealt{beers90})
of galaxy clusters as a function of $\left| v_{\rm rot} \right| /\sigma_p$, 
the ratio of absolute value of the rotation amplitude to the velocity dispersion (Left panel),
and as a function of $dv/dR (\equiv \sqrt{(\partial v/\partial X)^2 + (\partial v/\partial Y)^2})$,
the global velocity gradient (Right panel). Figure \ref{fig-rotcand} shows that there are four clusters (A1035, A1373, A2034, and A4053)
of which velocity dispersions are unusually large ($>$ 1300 km s$^{-1}$), while the majority of clusters are in the range of
290 km s$^{-1} < \sigma_{p} <$ 1050 km s$^{-1}$. Large velocity dispersions of these four clusters
might be related to the two superimposed or interacting clusters with slightly different systemic velocities (to be
discussed in Section \ref{notes}). 

It is expected that a rotating cluster has a large ratio of the absolute value of rotation amplitude to the velocity dispersion,
and has a large global velocity gradient across the cluster.
Therefore, among the selected 56 galaxy clusters, we selected 12 tentative rotating clusters of which
the ratios of the absolute value of rotation amplitude to the velocity dispersion are
greater than 0.53, and the global velocity gradients are greater than 380 km s$^{-1}$ Mpc$^{-1}$. 
We chose these critical values of $\left| v_{\rm rot} \right| /\sigma_p=0.53$ and $dv/dR=380$ km s$^{-1}$ Mpc$^{-1}$
because the rotation amplitudes of the clusters above these values are significant above $\sim2\sigma$ level.
In addition, the dips are seen at the critical values in both Figure \ref{fig-rotcand} (c) and (d), implying that
a segregation between the rotating and non-rotating clusters might exist, 
although it is not known whether there is a discrete boundary of the rotation amplitude 
between the rotating and non-rotating clusters or not.
These selection criteria are represented by dotted vertical lines in Figure \ref{fig-rotcand}.
In comparison, we plot the data of
Abell 2107 (star symbol) that is known to be a probable rotating cluster \citep{oh92,kalinkov05}.
It is found that Abell 2107 also satisfies our selection criteria for tentative rotating clusters. 

\end{enumerate}

Table \ref{tab-cand} lists the tentative rotating clusters with
Abell identification, RA and Dec (J2000), Bautz-Morgan (BM) type, the redshift derived in this study,
the velocity dispersion derived in this study, the number of member galaxies, X-ray luminosity,
and the reference for the X-ray luminosity. 
The sample of tentative rotating clusters are found from $z=0.028$ to $z=0.125$, and have 41$-$122 member galaxies.
There are seven clusters of which X-ray luminosities are estimated, and six clusters of which Bautz-Morgan types are III. 

The rotational properties derived using equation (\ref{eq-rot}) and (\ref{eq-fit}) are summarized
in Table \ref{tab-rot}. The first column gives the Abell identification. 
The second and the third columns represent, respectively, the position angle of rotation axis
and the rotation amplitude derived using equation (\ref{eq-rot}). 
The ratio of the absolute value of rotation amplitude to the velocity dispersion, $\left| v_{\rm rot} \right| /\sigma_p$,
is shown in the fourth column. The fifth column gives the position angle that is defined
by $\Theta_1$=tan$^{-1} \left[ (\partial v/\partial Y)/(\partial v/\partial X) \right] - \pi /2$. 
The final column gives the global velocity gradient derived using equation (\ref{eq-fit}).   
The uncertainties of these values represent $68\%$ $(1\sigma)$ confidence intervals.
We compute the uncertainties using 1000 artificial data set
constructed by choosing randomly cluster galaxies up to the number of cluster members in real data. 
The fitting procedure is done using 1000 trial data sets, 
the results are sorted, and the values corresponding to the 16th and the 84th percentiles are identified.
The uncertainties are defined as the offsets between these values and the values computed using the real data.
In Table \ref{tab-rot}, there are two clusters (A1035 and A1373) that have unusually large rotation amplitudes ($>$ 1300 km s$^{-1}$)
and velocity gradients ($>$ 1900 km s$^{-1}$ Mpc$^{-1}$) due to the superimposed two disparate groups (to be
discussed in Section \ref{notes}).
Other ten clusters show rotation amplitudes in the range of 190 km s$^{-1} < v_{\rm rot} < $ 830 km s $^{-1}$,
and velocity gradients in the range of 400 km s$^{-1}$ Mpc $^{-1} < dv/dR < $ 1080 km s $^{-1}$ Mpc $^{-1}$.

Figure \ref{fig-member} shows the plot of radial velocity versus clustercentric distance of galaxies
and the velocity distribution for 12 tentative rotating clusters.
In Figure \ref{fig-spvel}, we show the spatial distribution of cluster galaxies
with the measured velocities. Most clusters show that the galaxies of which velocities are greater than the systemic velocity of the
cluster (open circles) and the galaxies of which velocities are less than the systemic velocity of the
cluster (filled circles) are spatially separated well by the rotation axis (Y1 or Y2). Interestingly,
the position angles of the rotation axes, Y1 (determined from equation (\ref{eq-rot})) and Y2 (determined from equation (\ref{eq-fit})),
coincide well (the difference of position angles between Y1 and Y2 is less than 20 degree) except for three clusters (S0001, A1399, and A2169).

We plot the radial velocities as a function of
a projected distance along the direction perpendicular to the position angle ($\Theta_0$) of the rotation axis (Y1)
in Figure \ref{fig-xvel} which is available only in electronic issue.
The velocity gradients along the X1 axis are marginally seen in some clusters (A1035, A1373, A1474, and A2162),
while those are not clearly seen in the other clusters.

\section{Global Cluster Properties}
\subsection {Cluster Morphology}\label{morph}

In order to identify the connection between the cluster morphology and the cluster dynamics,
it is useful to determine the ellipticity and the orientation of a cluster
using a spatial distribution of the selected member galaxies.
To determine the cluster shape, we use the method of the dispersion ellipse of the
bivariate normal frequency function of position vectors (e.g. \citealt{trumpler53,carter80,burgett04}).
The dispersion ellipse is defined in \citet{trumpler53} as the contour at which the density is 0.61 times the maximum density
of a set of points distributed normally with respect to two correlated variables, although
the points need not be distributed normally in order to determine the proper cluster shape.
From the first five moments of the spatial distribution,

\alpheqn
\begin{equation}
\mu_{10} = \frac{1}{N}\sum_{i=1}^N X_i
\end{equation}
\begin{equation}
\mu_{01} = \frac{1}{N}\sum_{i=1}^N Y_i
\end{equation}
\begin{equation}
\mu_{20} = \frac{1}{N}\sum_{i=1}^N X^2_i - \left( \frac{1}{N}\sum_{i=1}^N X_i \right)^2
\end{equation}
\begin{equation}
\mu_{11} = \frac{1}{N}\sum_{i=1}^N X_iY_i - \frac{1}{N^2}\sum_{i=1}^N X_i\sum_{i=1}^N Y_i 
\end{equation}
\begin{equation}
\mu_{02} = \frac{1}{N}\sum_{i=1}^N Y^2_i - \left( \frac{1}{N}\sum_{i=1}^N Y_i \right)^2 \; \; \mbox{,}
\end{equation}
\reseteqn

where X and Y are clustercentric distances in the direction of right ascension
and declination, respectively. The semi-major and minor axes of the ellipse, $\Gamma_A$ and $\Gamma_B$,
are derived by solving the equation of
\begin{equation}
\left| \begin{array}{cc}
\mu_{20}-\Gamma^2 & \mu_{11} \\
\mu_{11}     & \mu_{02}-\Gamma^2 \\
\end{array} \right|
= 0 \; \; \mbox{.}
\end{equation}

The position angle of the major axis measured from north to east is given by
\begin{equation}
\Theta_2 = \cot^{-1}\left( -\frac{\mu_{02} - \Gamma_A^2}{\mu_{11}}\right) + \frac{\pi}{2} \; \; \mbox{,}
\end{equation}
and the ellipticity is defined by
\begin{equation}
\epsilon = 1 - \frac{\Gamma_B}{\Gamma_A} \; \; \mbox{.}
\end{equation}

The major/minor axis, position angle, and ellipticity 
for our tentative rotating clusters that we derived 
are listed in Table \ref{tab-mor}. Interestingly, the ellipticities of S0001 ($\epsilon=$0.52) and A1399 ($\epsilon=$0.57) are
unusually larger than those of other clusters. However, the ellipticity of S0001 might be uncertain because no data are available for 
$\sim30\arcmin$ away to the North from the S0001 center as seen in Figure \ref{fig-spvel}. 

\subsection {Analysis of Substructure}

The analysis of substructure is a useful diagnostic tool for 
understanding of the dynamical state of galaxy clusters.
A useful discussion for several substructure tests is
given in \citet{pinkney96}.
We have derived a number density
map using the spatial position (2D) of member galaxies,
and performed 1D and 3D substructure tests for the sample of tentative rotating clusters.  

The majority of 1D (velocity histogram) substructure tests are normality
tests. We present the results of five 1D tests for tentative rotating clusters
in Table \ref{tab-sub}.
The values of I test \citep{teague90} are shown in the first and the second column. $I_{90}$ is the
critical value for rejecting the Gaussian hypothesis at 90 \% confidence level. Therefore,
a velocity distribution is considered to be a non-Gaussian if $I>I_{90}$.
It is found that all clusters satisfy the Gaussian hypothesis using the I test except for S1171 and A4053.

The skewness that is a measure of the degree of asymmetry of a distribution, and its
confidence level of rejecting normality are given in the third and the fourth columns. 
Positive (Negative) skewness indicates that the distribution is skewed to the right (left), with a longer tail
to the right (left) of the distribution maximum. 
The kurtosis that is the degree of peakedness of a distribution and its
confidence level of rejecting normality are given in the fifth and the sixth columns.
Positive values of the kurtosis indicate pointed or peaked distributions,
while negative values flattened or non-peaked distributions.
The skewness test rejects a Gaussian distribution with a confidence level of $>99\%$ for only S1171 and A4053,
being consistent with the result of I test. The kurtosis test rejects a Gaussian hypothesis 
of A1035, A1373, A1474, and A2162 with a confidence level of $>95\%$.

From the seventh to the tenth column, we present the Asymmetry Index (AI) and the Tail Index (TI) 
introduced by \citet{bb93} with their confidence levels.
The AI measures the symmetry in a population by comparing gaps in the data on the
left and right sides of the sample median, and TI compares the spread of the data at 90 \%
level to the spread at the 75 \% level.
The Gaussian hypothesis for A1035 and A1474 are rejected by both AI and TI tests with a confidence level
of $>95\%$. 

\begin{figure}
\includegraphics [width=145mm] {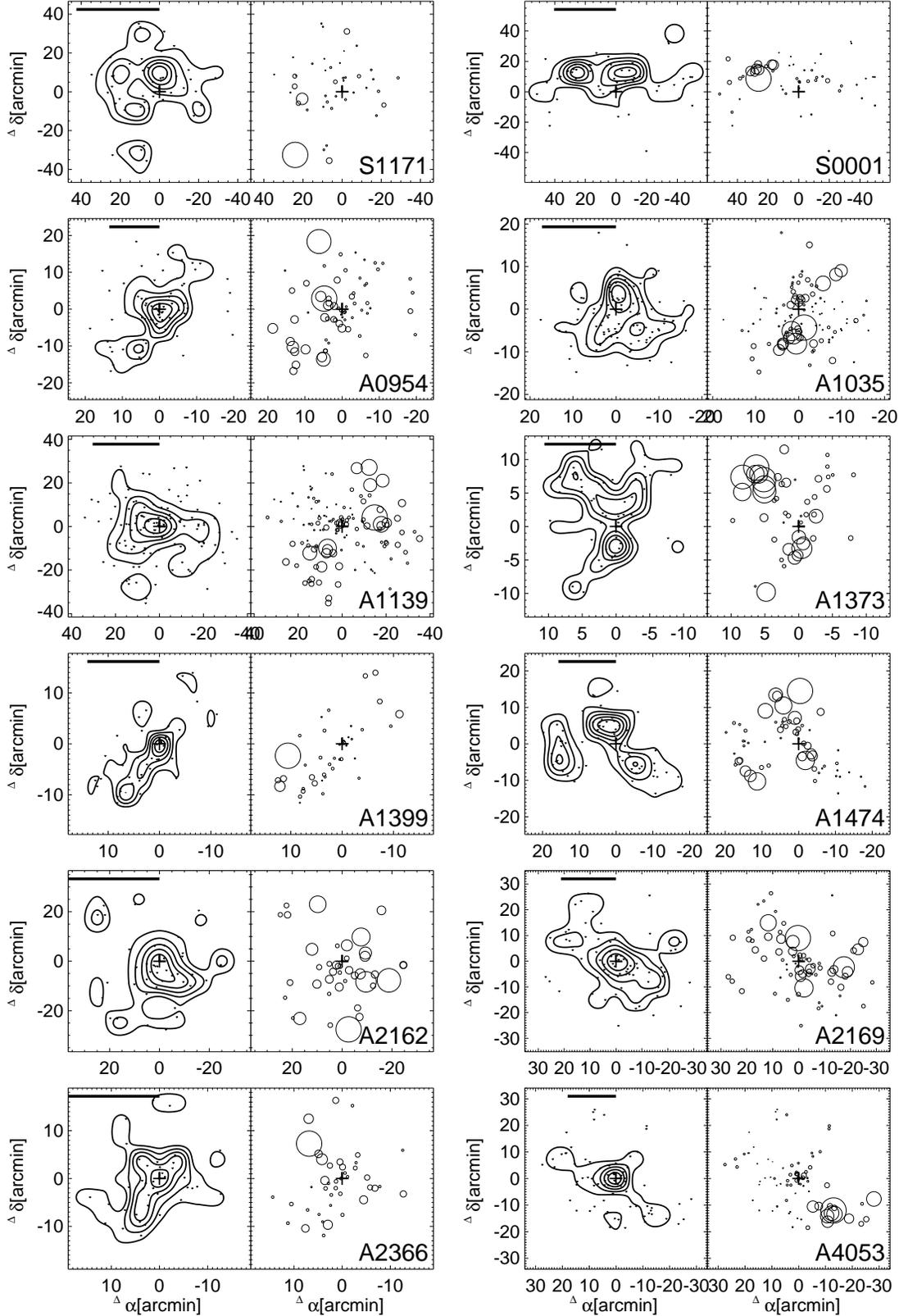}
\caption{{\it Left}: Galaxy number density maps for tentative rotating clusters. 
The member galaxies are represented by dots, and the number density
contours are overlaid. The cross indicates the cluster centroid, and the thick horizontal bar represents the
physical size of 1 Mpc.
{\it Right}: Dressler-Shectman (DS) plots for the same clusters. Each galaxy is plotted by a circle with
the diameter proportional to $e^\delta$.
}\label{fig-sub}
\end{figure}

We have constructed the number density contours for the clusters
using a different bin size of $0.25 R \times 0.25 R ~\rm{(Mpc)^2}$ depending on
the cluster size ($R=4\sqrt{\Gamma_A \Gamma_B}/1.5$).
$\Gamma_A$ and $\Gamma_B$ are used in the unit of Mpc, 4 is an arbitrary constant, 
and 1.5 (Mpc) is normalizing constant. We have smoothed the contours
using a cubic convolution interpolation method. The contour interval,
(the maximum density in a cluster)$/6$, is also
determined according to the maximum number density of the clusters. 
We plot the number density map in the first and the third columns in Figure \ref{fig-sub}.

Using the velocity data and positional information of galaxies, we have performed $\Delta$ test \citep{ds88}
that computes the local deviations from the systemic velocity ($v_{\rm sys}$) and dispersion ($\sigma_p$)
of the entire cluster. For each galaxy, the deviation is defined by

\begin{equation}
\delta^2 =  \frac{N_{nn}}{\sigma_p^2} \left[ (v_{\rm local}-v_{\rm sys})^2 +
(\sigma_{\rm local}-\sigma_p)^2 \right] \; \; \mbox{,}
\end{equation}

where $N_{nn}$ is the number of galaxies that defines the local environment, taken
to be $\sim \sqrt{N_{gal}}$ in this study. The sum of $\delta$ over all cluster galaxies
in a cluster, $\Delta$, 
is used to quantify the presence of substructures.
It is approximately equal to the total number
of galaxies in a cluster in the case of no substructure, while it is larger 
in the presence of substructure.

The statistical significance of the deviation is computed by Monte Carlo
simulation. The velocities are randomly
assigned to the galaxies at the observed galaxy positions, and $\Delta_{\rm sim}$ is
computed for each simulated cluster. We construct 1000 simulated clusters and
compute $\Delta_{\rm sim}$ for each simulation. We present $\Delta_{\rm obs}$
which is computed using real data and the fraction of simulated clusters with 
$\Delta_{\rm sim}>\Delta_{\rm obs}$ in the final two columns of Table \ref{tab-sub}.
Small values of $f(\Delta_{\rm sim}>\Delta_{\rm obs})$ indicate the statistically
significant substructure in clusters. Most clusters have small values of $f(\Delta_{\rm sim}>\Delta_{\rm obs})$,
indicating significant substructures or spatially correlated velocities of galaxies except for S1171, A2162, and A2366.

We plot the positions of cluster galaxies represented by open circles of which
radii are proportional to $e^{\delta}$ in the second and the fourth columns of Figure \ref{fig-sub}.
A large circle denotes a galaxy which is deviant in either velocity or
dispersion compared with nearby galaxies, therefore groups of large circles indicate
the presence of substructure. No substructures of S1171, A2162, and A2366 were found 
in $f(\Delta_{\rm sim}>\Delta_{\rm obs})$ test, as confirmed in Figure \ref{fig-sub}.

\section{Detailed analysis of individual clusters}\label{notes}

A single cluster in rotation is expected to show a spatial segregation of high and low
velocity galaxies, and to have a single number density peak unlike the case two clusters
are still in the middle of merging. Therefore, in the sample of tentative rotating clusters,
we select as probable rotating clusters the galaxy clusters that show a spatial segregation of high and low velocity galaxies in the
spatial-density map of Figure \ref{fig-spvel}, and show a single peak in the number density map of Figure \ref{fig-sub}.
We describe the selection procedure of the probable rotating cluster for an individual cluster in this section.

\subsection{Abell S1171}
Abell S1177 is the nearest ($z\sim0.028$) cluster in the sample of tentative rotating clusters. 
In the plot of velocity versus clustercentric distance of member galaxies (Figure \ref{fig-member}),
it is seen that some galaxies have large velocity deviations ($cz-\overline{cz}\sim$1400 km s$^{-1}$) 
from the cluster main body ($\overline{cz}\sim$8400 km s$^{-1}$).
If we consider only the galaxies in the cluster main body rejecting these galaxies ($cz>$9500 km s$^{-1}$) as interlopers 
(in order to test the effect of this sub-component),
we obtain $|v_{rot}|$/${\sigma}_{p}$=0.33 and $dv/dR$=299 km s$^{-1}$ Mpc$^{-1}$. 
Then this cluster is not included in the the sample of tentative rotating clusters.

\subsection{Abell S0001}
\begin{figure}
\includegraphics [width=165mm] {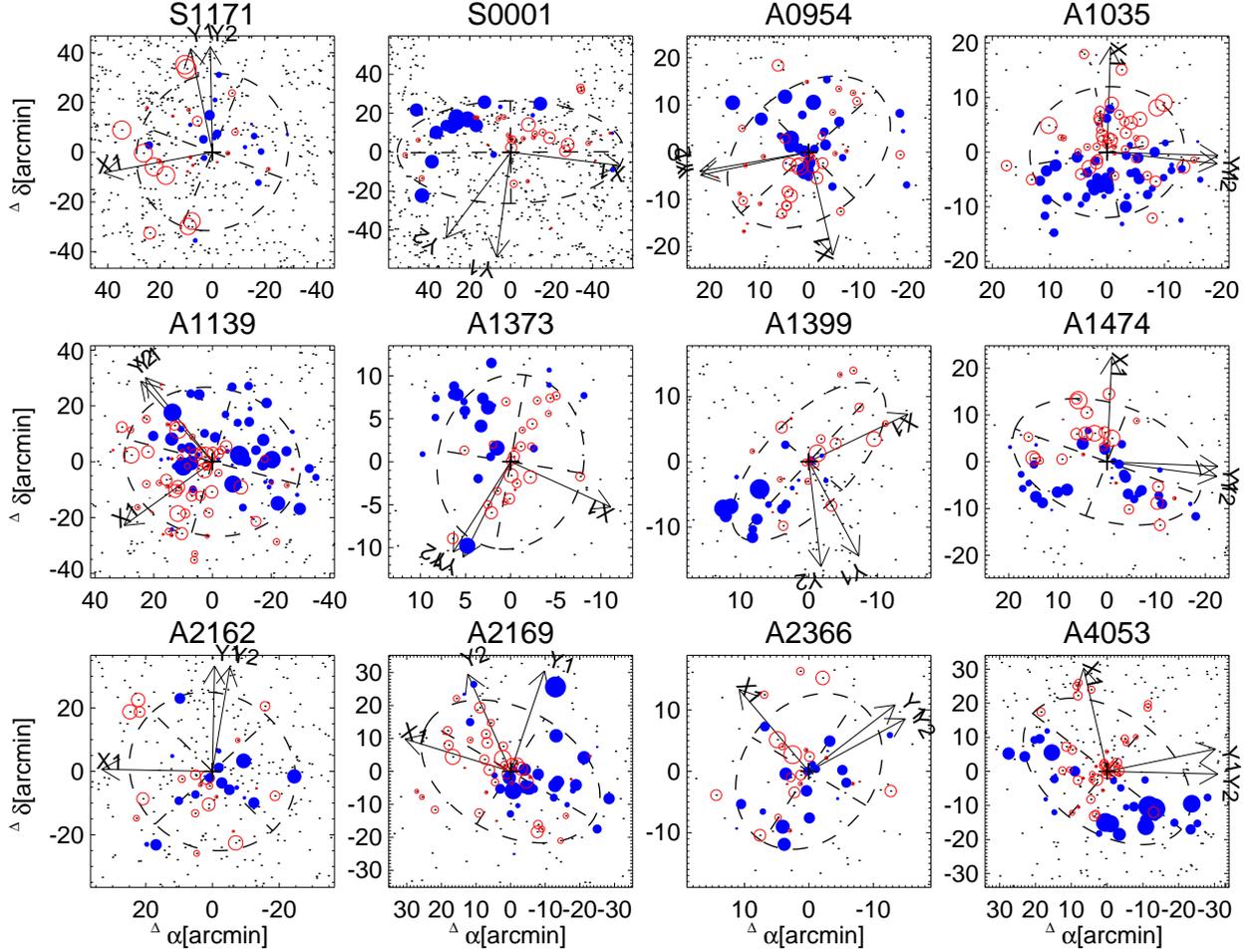}
\caption{Spatial distribution of cluster galaxies with measured velocities in 12 tentative rotating clusters.
The cluster galaxies of which velocities are greater than
the systemic velocity of their cluster are plotted
by open symbols, while those of which velocities are less than
the systemic velocity by filled symbols. The symbol size is
proportional to the velocity deviation. The observed galaxies
are represented by dots. 
The position angles, $\Theta_0$ and $\Theta_1$, derived using equation (\ref{eq-rot}) and (\ref{eq-fit})
are represented by arrows (axis Y1 and Y2).
The perpendicular axis to the position angle of rotation axis (Y1) is shown by X1.
The dashed ellipse indicates the twice enlarged dispersion ellipse, and two dashed lines denote the
major and minor axis of the dispersion ellipse, respectively.
}\label{fig-spvel}
\end{figure}

\begin{figure}
\includegraphics [width=165mm] {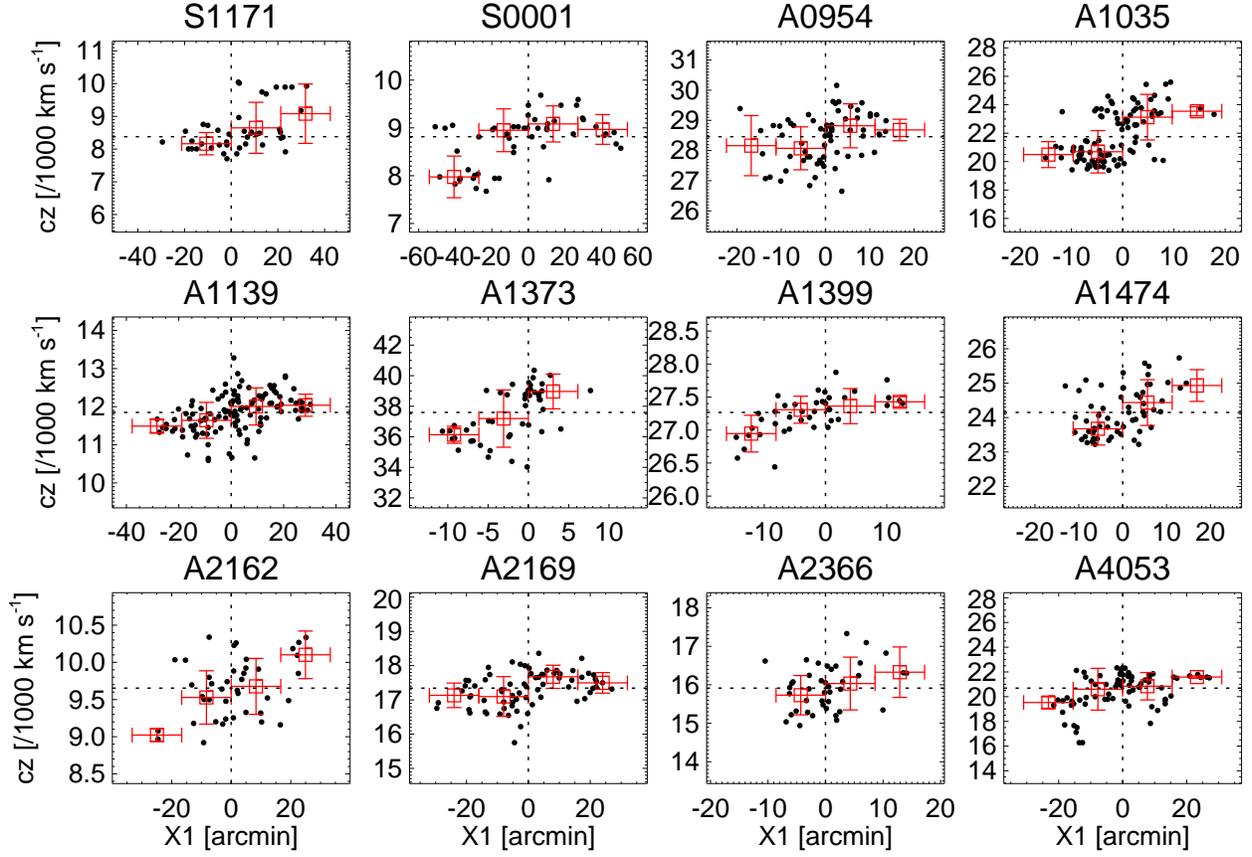}
\caption{Radial velocities of cluster galaxies as a function of
projected distance along axis X1 shown in Figure \ref{fig-spvel}.
Large open squares indicate the mean radial velocity of cluster galaxies 
in a distance bin that is represented by a horizontal errorbar. The vertical
errorbar denotes the velocity dispersion of galaxies in a distance bin.
}\label{fig-xvel}
\end{figure}

Abell S0001 shows a low velocity tail ($cz\sim$7800 km s$^{-1}$) in the velocity histogram of Figure \ref{fig-member}.
These galaxies in the low velocity tail are spatially concentrated, appear as a substructure (to the North-East) in Figure \ref{fig-sub}.
The number density map shows two local density maxima which coinside with the positions of substructures.
Therefore, the large amplitudes of rotation and velocity gradient are considered to be caused by two separated subclusterings.
This cluster shows a large ellipticity, and that the position angle of rotation axis (Y1) shown in Figure \ref{fig-spvel} is similar to that of cluster minor axis ($\Gamma_B$). However, the determination of cluster morphology is uncertain, since no data are available for 
$\sim30\arcmin$ away to the North from the cluster center (see Figure \ref{fig-spvel}).

\subsection{Abell 0954}
Abell 0954 is surveyed in both SDSS and 2dFGRS, and is one of the most X-ray luminous clusters in the sample of tentative rotating clusters.
In Figure \ref{fig-spvel}, the position angles of rotation axes 
determined by the rotation fit (Y1) and the global velocity gradient fit (Y2) coincide well,
but those position angles appear not to be correlated with the position angle of major or minor axis of the dispersion ellipse.
The results of 1D substructure tests in Table \ref{tab-sub} show that the velocity distribution does not deviate from Gaussian significantly.
The number density map and Dressler-Shectman (DS) plot in Figure \ref{fig-sub} show a weak sign of subclustering $\sim14\arcmin$ away 
to the South-East,
and it is suspected that this subclustering might cause the large amplitude of rotation. 
In order to test the effect of the SE subclustering, we fit the velocities
of member galaxies that are located within 10$\arcmin$ from the cluster center using equation (\ref{eq-rot}) and (\ref{eq-fit}). The fitting results in $|v_{rot}|$/${\sigma}_{p}$=0.54 and $dv/dR$=948 km s$^{-1}$ Mpc$^{-1}$ 
with the same position angles of rotation axes as the results based on all member galaxies.
This implies that the rotation and the velocity gradient are generated in the core region with a single density peak.
Therefore we select this cluster as a probable rotating cluster.

\subsection{Abell 1035}
\begin{figure}
\includegraphics [width=165mm] {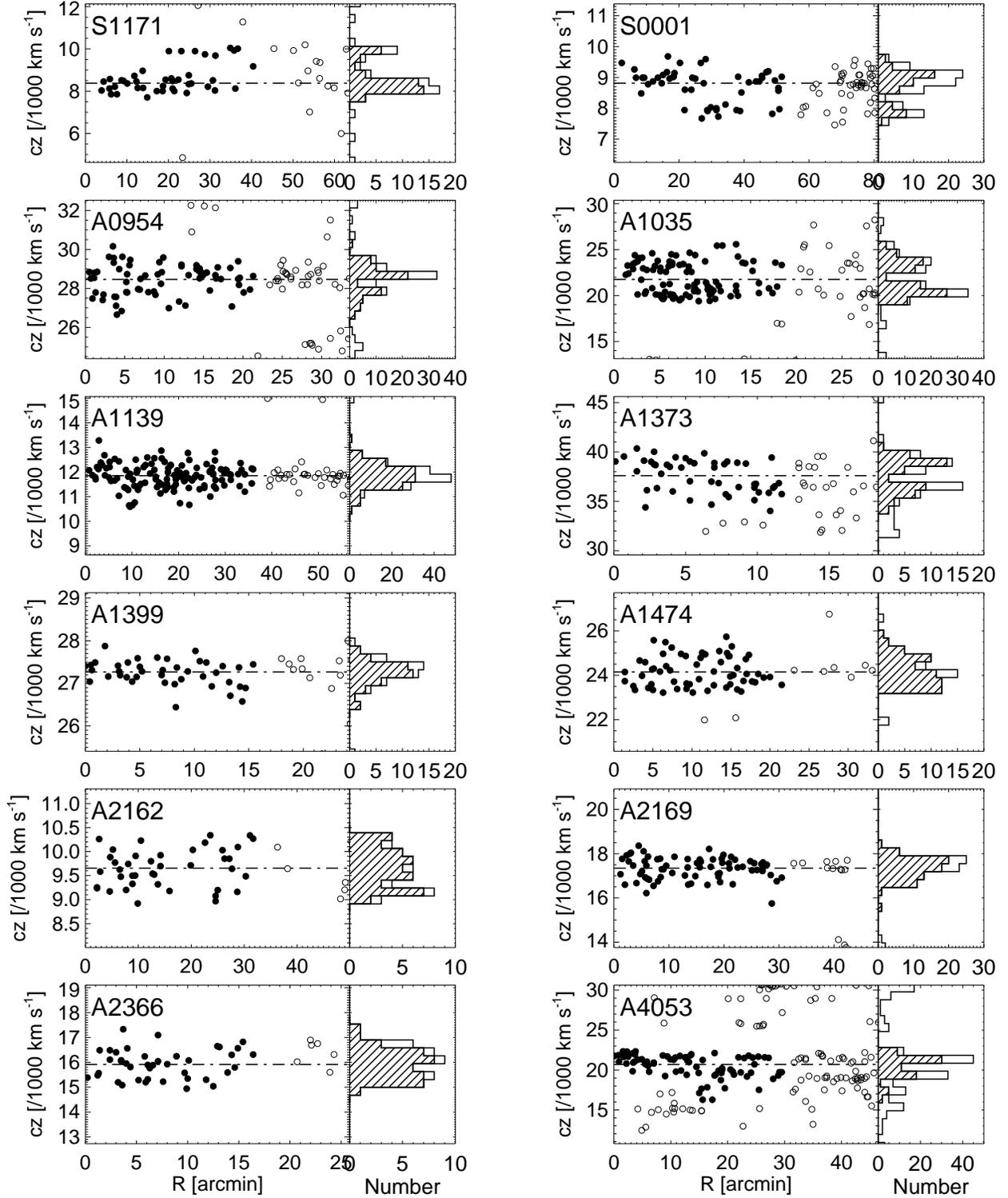}
\caption{Radial velocity versus clustercentric distance of galaxies and
the velocity distribution for 12 tentative rotating clusters. Filled circles indicate the galaxies selected
as cluster members, while open circles the galaxies not selected as cluster members.\
The horizontal dot-dashed line indicates the systemic velocity of the clusters determined in Table \ref{tab-cand}.
The velocity distribution for the member galaxies is shown by a shaded histogram and
that for the entire observed galaxies is shown by an open histogram. 
} \label{fig-member}
\end{figure}

Abell 1035 has the largest X-ray luminosity and the velocity dispersion in the sample of tentative rotating clusters.
While \citet{flin06} found no substructure using only positional information of galaxies by applying wavelet analysis,
\citet{Maurogordato97} found a bimodal velocity distribution for this cluster using the velocity data of 19 member galaxies, 
as confirmed in Figure \ref{fig-member} based on 97 member galaxies in this study.
Our result in Figure \ref{fig-member} implies that substructures may exist.
In Figure \ref{fig-spvel}, the galaxies of which velocities are greater than the systemic velocity of the
cluster (open circles) and the galaxies of which velocities are less than the systemic velocity of the
cluster (filled circles) are spatially separated clearly by the rotation axis (Y1 and Y2). In addition,
the DS plot in Figure \ref{fig-sub} shows two subclusterings (North and South) that coinside with
high density regions in the galaxy number density map. Therefore, the large amplitudes of 
the rotation and the velocity gradient including the large velocity dispersion (listed in Table \ref{tab-cand}),
can be explained by two superimposed or interacting clusters with slightly different systemic velocities.

\citet{as06} found that the spin-vector orientations of the galaxies in this cluster tend to lie parallel to the
Local Supercluster (LSC) plane, and the spin-vector projections of the galaxies tend to be oriented perpendicular with respect
to the direction of the LSC center.
In this study, we found that the angle between the rotation axis (Y1) of this cluster and the LSC plane is $56^{+11}_{-9}$ deg.
Since the angle between the spin-vector orientations of the galaxies in this cluster and the LSC plane was found to be $<45$ deg in \citet{as06},
it appears that the spin-vector orientations of the galaxies determined in \citet{as06}
is not related to the rotation axis of this cluster determined in this study.

\subsection{Abell 1139}
Abell 1139 is surveyed in both SDSS and 2dFGRS, and has the largest number of member galaxies in our sample.
A global velocity gradient for this cluster is identified by \citet{burgett04} using 2dFGRS data with 106 member galaxies.
We confirm the global velocity gradient or the rotation using 122 member galaxies with an aid of SDSS data (see Figrue \ref{fig-xvel}).
The DS plot in Figure \ref{fig-sub} shows a presence of strong substructures 
to the North-West and to the South-East, while the velocity histogram shows a unimodal distribution. 
It is known that positional information of galaxies only shows no clear substructure for this cluster
(e.g. \citealt{Krywult99,flin06}), but a distinguishable substructure is found based on 
both positional information and velocity data (e.g. \citealt{burgett04}).
As seen in the spatial-velocity plot in Figure \ref{fig-spvel}, the substructures
are due to the galaxies with different radial velocities. Interestingly, unlike Abell 1035, the galaxy number density map shows 
a single density peak near the cluster center rather than strong concentrations around two substructures. This may indicate
a single cluster in rotation or two overlapping clusters that either merge or depart from each other as pointed in \citet{burgett04}. 
Therefore, we select this cluster as a probable rotating cluster.

\subsection{Abell 1373}
Abell 1373 is the most distant cluster ($z\sim0.125$) in our sample of tentative rotating clusters, and is surveyed in both SDSS and 2dFGRS.
The velocity distribution is bimodal, and the galaxies with high velocities ($cz\sim$39000 km s$^{-1}$) and the galaxies
with low velocities ($cz\sim$36000 km s$^{-1}$) are spatially separated by the rotation axis (see Figure \ref{fig-spvel}).
Similar to the case of Abell 1035, the substructures introduced by the clustering of galaxies with different velocities coincide
with the galaxy number density peaks, implying that the large amplitudes of rotation and velocity gradient are introduced by
two disparate groups (see Figure \ref{fig-sub}). 

\subsection{Abell 1399}
Abell 1399 is one of the member clusters in the Leo A Supercluster \citep{einasto01}.
This cluster has the smallest velocity dispersion in our sample of tentative rotating clusters, and is surveyed in both SDSS and 2dFGRS.
The velocity distribution is unimodal, but the spatial-velocity plot of Figure \ref{fig-spvel} shows a segregation 
of galaxies with different velocities.
Intriguingly, the dispersion ellipse is significantly flatted with the ellipticity of 0.57. Since the position angles 
(138 deg for the major axis and 48 deg for the minor axis) of the dispersion ellipse
deviate from those (207 deg for Y1 and 186 deg for Y2) of the rotation axes, it appears that
the high ellipticity is not related to the global rotation of this cluster. Figure \ref{fig-sub} shows that there is no strong
sign of the correlation between the number density peaks and the substructures.
Therefore, we select this cluster as a probable rotating cluster.

\subsection{Abell 1474}
Abell 1474 is one of the member clusters in the Virgo-Coma Supercluster \citep{einasto01}.
\citet{flin06} found no substructure using only positional information of galaxies, but the 1D substructure tests
indicate that the velocity distribution deviates from the Gaussian significantly. In Figure \ref{fig-sub}, 
the DS plot shows that there are three substructures (to the North-East, to the South-West, and to the South-East) 
that coincide with three number density peaks in the left panel. It indicates that the large amplitudes of rotation and 
velocity gradient are due to the different velocities of sub-groups.

\subsection{Abell 2162}
Abell 2162 is one of the nearest clusters in our sample, and is a member cluster in the Hercules Supercluster \citep{einasto01}.
Neither a distinct substructure nor a significant local density peak is found except for the cluster center in Figure \ref{fig-sub}.
Interestingly, it appears that the radial distribution of the member galaxies in Figure \ref{fig-member} is discontinuous
at $\sim 18 \arcmin$. Therefore, if we compute the rotation amplitude and 
the global velocity gradient using only the galaxies at the inner region ($< 18 \arcmin$),
we obtain $|v_{rot}|$/${\sigma}_{p}$=0.64 and $dv/dR$=635 km s$^{-1}$ Mpc$^{-1}$, still satisfying the criteria for the rotating clusters.
Therefore, we select this cluster as a probable rotating cluster. 

\subsection{Abell 2169}
In Figure \ref{fig-spvel}, the galaxies of which velocities are greater than the systemic velocity of Abell 2169 (open circles) and 
the galaxies of which velocities are less than the systemic velocity (filled circles), are separated well by the rotation axis (Y1). The
position angle (342 deg) of rotation axis (Y1) is similar to that (328 deg) of the minor axis of the dispersion ellipse,
implying that the flatness may be caused by the cluster rotation. The substructures (North-East and South-West) seen in the DS plot,
are separated by the cluster centroid that coincides with the single number density peak. Therefore, this cluster is selected as a probable
rotating cluster. 

\subsection{Abell 2366}
\citet{jf99} classified the morphology of Abell 2366 using {\it Einstein} X-ray images 
as a ``single'', means a cluster in which no substructures or no departures from symmetry
are found. \citet{flin06} also found no substructure using only positional information of galaxies 
by applying wavelet analysis. Similarly, the galaxy number density map and the DS plot in Figure \ref{fig-sub} indicate
no strong evidence of substructures. Therefore, we select this cluster as a probable rotating cluster.

\subsection{Abell 4053}
Abell 4053 is one of the member clusters in the Pisces-Cetus Supercluster \citep{pr05}, 
and is known to be contaminated by a foreground group at $z=0.0501$ \citep{mazure96}.
The galaxies in the foreground group ($cz\sim$15000 km s$^{-1}$) are seen in Figure \ref{fig-member}, 
and are rejected from determining the membership of Abell 4053. 
However, the galaxies at the low velocity tail ($cz\sim$17500 km s$^{-1}$) are selected as members of Abell 4053. 
These galaxies are found in the South-West region from the cluster centroid (see Figure \ref{fig-spvel}), and
are detected as a subcluster in the DS plot.  
In order to estimate the effect of these galaxies in the low velocity tail, we determine the amplitudes of the rotation and the velocity
gradient using only the galaxies in the main body ($cz>$18000 km s$^{-1}$). The fitting results in 
$|v_{rot}|$/${\sigma}_{p}$=0.47 and $dv/dR$=716 km s$^{-1}$ Mpc$^{-1}$, which does not satisfy the criteria for the tentative rotating cluster.
Therefore, this cluster is not selected as a probable rotating cluster.

We summarize the global kinematic properties for 12 tentative rotating clusters in Table \ref{tab-sum}.
The column (1) gives the Abell identification. The column (2), (3) and (4) represent 
the existence of substructure resulted from 1D, 2D and 3D tests, respectively. The cluster morphology determined in Section \ref{morph}
is given in the column (5): `Spherical' for ellipticity $<0.2$ and `Elongated' for ellipticity $\ge0.2$.
The dynamical status determined in Section \ref{discuss} is given in the column (6) and
the selection of probable rotating clusters is in the final column (7).

\section{Discussion}\label{discuss}
\subsection{Dynamical status of probable rotating clusters}

\begin{figure}
\includegraphics [width=145mm] {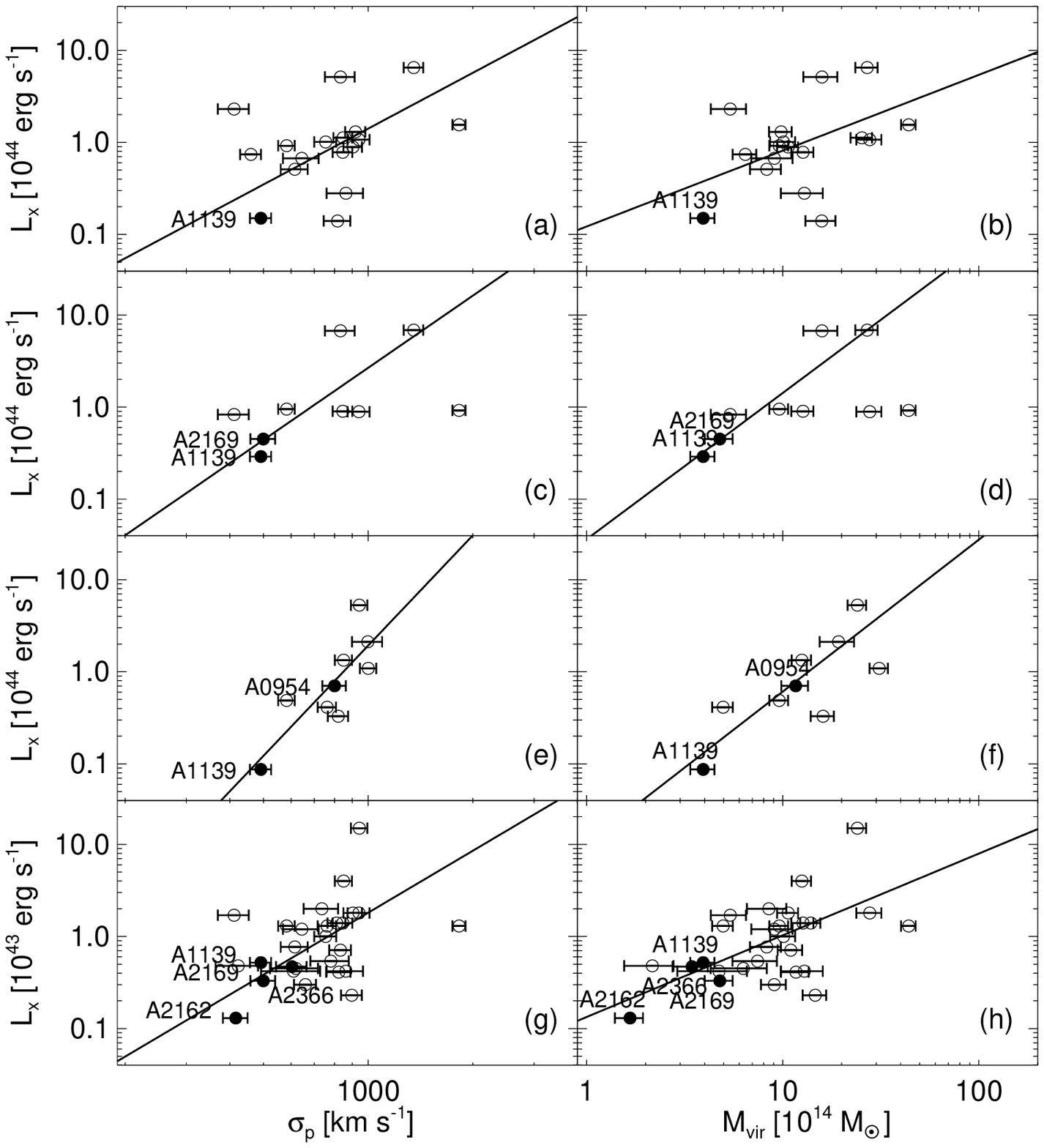}
\caption{X-ray luminosity $L_{\rm x}$ (0.1$-$2.4 keV) versus the velocity dispersion and the virial mass for 
the probable rotating clusters (filled circles) in comparison with the non-rotating galaxy clusters (open circles)
of the selected 56 galaxy clusters. X-ray luminosities in (a) and (b) are from \citet{bohringer00},
those in (c) and (d) from \citet{ebeling98,ebeling00}, those in (e) and (f) from \citet{bohringer04},
and those (0.5$-$2.0 keV) in (g) and (h) from \citet{ledlow03}. The solid line indicates the best fit for each panel.
}\label{fig-xray}
\end{figure}

The study of dynamical state of the propable rotating clusters is useful for understanding the origin of the global rotation across the cluster.
It is expected from self-similar models that a relationship between the X-ray luminosity and the velocity dispersion of the clusters exists
as $L_{\rm x}\propto \sigma_p^4$ (e.g. \citealt{qm82}). For galaxy clusters in SDSS, \citet{popesso05} found the relation of
$L_{\rm x}\propto \sigma_p^{3.68\pm0.25}$. For galaxy clusters in 2dFGRS, \citet{hilton05} found 
the relation of $L_{\rm x}\propto \sigma_p^{4.8\pm0.7}$, and suggested that high-$L_{\rm x}$ clusters are more dynamically evolved systems than the low-$L_{\rm x}$ clusters. 
In Figure \ref{fig-xray}, we plot the X-ray luminosity as a function of the velocity dispersion and the virial mass for the probable
rotating clusters (filled circles) of which X-ray luminosities are available in the literature in comparison 
with the non-rotating galaxy clusters (open circles) 
of which X-ray luminosities are available in the literature among the selected 56 galaxy clusters.
We plot the X-ray luminosities from the several literature separately. 
It appears that most galaxy clusters shown in Figure \ref{fig-xray} follow $L_{\rm x} - \sigma_p$ and $L_{\rm x} - M_{vir}$ relations (solid lines)
determined in this study.
Moreover, several interesting features are seen in the probable rotating clusters.
First, the probable rotating clusters do not deviate from the best-fit line, implying that those clusters
are in dynamical equilibrium. Secondly, the velocity dispersions and the virial masses for the probable rotating clusters
appear to be smaller than those for the non-rotating clusters. In addition, the X-ray luminosities for the probable rotating clusters
also appear to be smaller than those for the non-rotating clusters. If we adopt the result of \citet{hilton05}, the probable rotating
clusters that have low X-ray luminosities might have a lower fraction of early-type, passively evolving galaxies than
the galaxy clusters with high X-ray luminosities. 

If galaxy clusters rotate, the velocity dispersions and the corresponding virial masses should be corrected for the cluster rotation
(e.g. \citealt{kalinkov05}). 
We derived the velocity dispersions about the best fit rotation curve of equation (\ref{eq-rot}) and
the corresponding corrected virial masses for the probable rotating clusters. The results are summarized in Table \ref{tab-corr}.
The first and the second columns represent, respectively, the velocity dispersion
about the mean velocity of the cluster and that about the best fit rotation curve of equation (\ref{eq-rot}).
The third and the fourth columns represent, respectively, the virial mass
using the rotation uncorrected (the first column) velocity dispersion and that using the rotation corrected (the second column) velocity dispersion.
The velocity dispersions are reduced by 6$-$14\%, and on the average are reduced by 11\%.
The corresponding virial masses are also reduced by 11$-$27\%, and on the average are reduced by 21\%.

The dynamics of the brightest cluster galaxies (BCGs) or cD galaxies in galaxy clusters are also useful for understanding the
formation history of galaxy clusters (e.g. \citealt{oh01}). In particular, a peculiar velocity of BCGs, which is defined
by $v_p = v_{BCG}-v_{cl}$, where $v_{BCG}$ is a radial velocity of BCGs and $v_{cl}$ is a mean velocity of the cluster,
is a useful indicator of the dynamical state for the cluster \citep{oh01}.
To estimate the peculiar velocities of BCGs in clusters,
we first identified the BCGs that have the smallest
$b_{\rm J}$ magnitudes in the catalog of member galaxies for the selected 56 galaxy clusters. 
Then we conducted a visual inspection of cluster images 
whether there are any brighter galaxies than the selected BCGs using the catalog of member galaxies.
Since some brightest galaxies in galaxy clusters were not covered due to the observational difficulties 
such as a fiber collision and a saturation, 
we finally selected 20 galaxy clusters of which BCGs in the catalog agree with those in the images.
Using the sample of selected 20 galaxy clusters, we plot the absolute value of the peculiar velocity of BCGs as a function
of the clustercentric distance, the redshift of clusters as a function of the absolute magnitude of BCGs in 
$b_{\rm J}$ band, and the velocity dispersion of clusters as a function of the absolute value of the peculiar velocity of BCGs.
It is seen that the absolute values of peculiar velocities of the BCGs are in the range of 10$-$780 km s$^{-1}$, and the median value
of peculiar velocities is 257 km s$^{-1}$. The mean uncertainty for the values of peculiar velocities is
168 km s$^{-1}$. In addition, the clustercentric distances of BCGs are in the rage of 0$-$300 kpc, and
the median value of clustercentric distances is 82 kpc. Interestingly, the peculiar velocities and the clustercentric distances
of BCGs for two (A0954 and A1399) probable rotating clusters are smaller than the median value of each parameter. The small peculiar velocity
and the small spatial deviation of BCGs from the cluster center indicate that the clusters are in the dynamical equilibrium.
It may imply that the clusters do not undergo a recent merging if the clusters are formed through a repeated merging.
In addition, the absolute magnitudes of BCGs for two probable rotating clusters do not show any significant deviation from the distribution of the
non-rotating galaxy clusters. 

\subsection{Probable rotating clusters and spatial orientations of cluster galaxies}
\begin{figure}
\includegraphics [width=145mm] {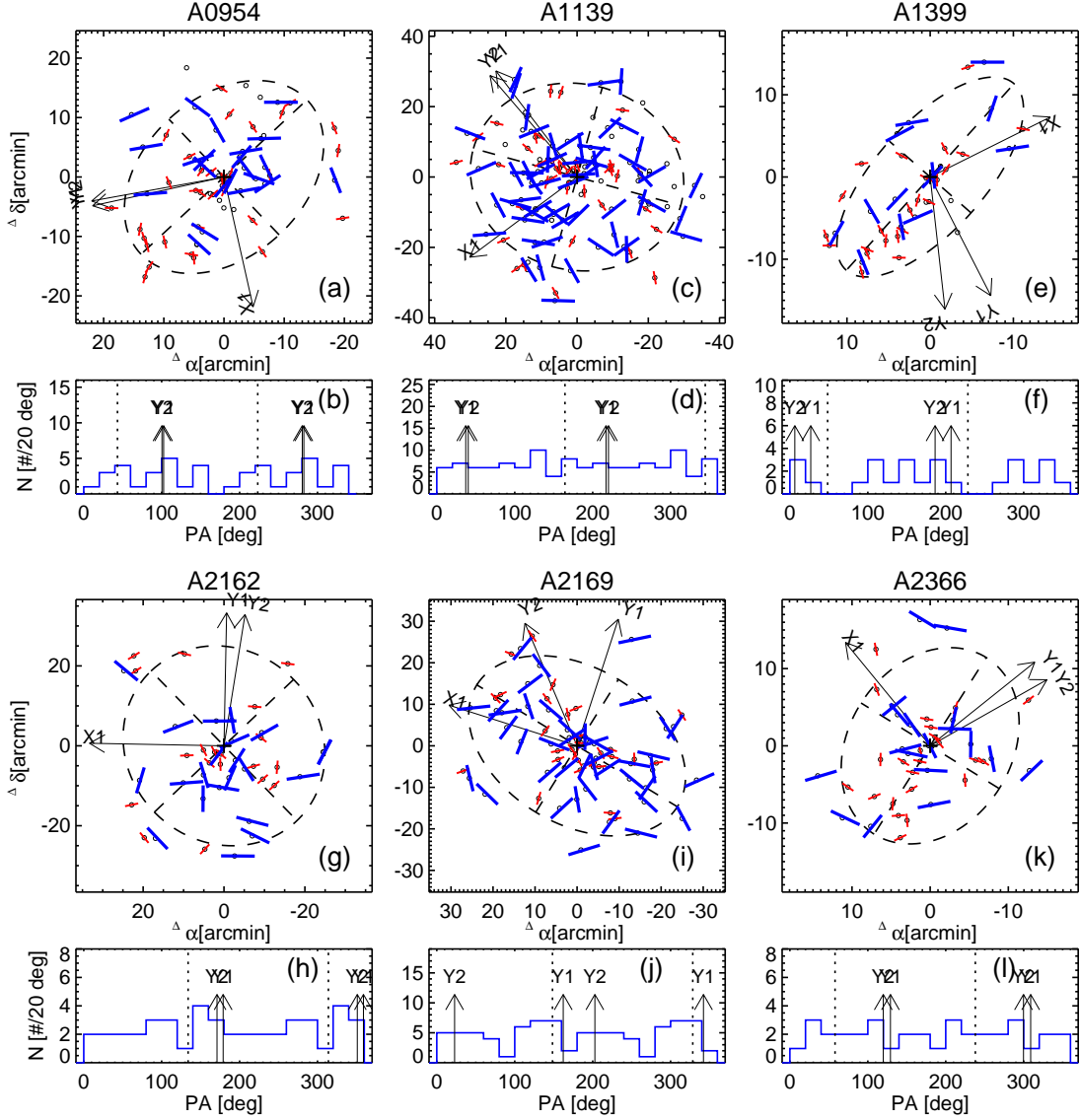}
\caption{(a, c, e, g, i, k) Spatial distribution for the projected spin-vector orientations of cluster galaxies in six probable rotating clusters.
The cluster galaxies are plotted by open circles. Long solid lines represent the projected spin-vector orientations for the moderately edge-on galaxies
(axial ratio is less than 0.72 in SDSS and eccentricity is equal to or greater than 0.33 in 2dFGRS), while
short solid lines represent those for the face-on galaxies
(axial ratio is equal to or greater than 0.72 in SDSS and eccentricity is less than 0.33 in 2dFGRS).
The galaxies with unknown orientations are plotted by only open circles.
The dispersion ellipse and the rotation axes (Y1 and Y2) shown in Figure \ref{fig-spvel} are overlaid.
(b, d, f, h, j, l) Histogram for the projected spin-vector orientations of cluster galaxies.
We present the histogram for only the moderately edge-on galaxies.
The rotation axes (Y1 and Y2) are represented by the vertical arrows and the minor axis of the dispersion ellipse
is represented by the vertical dotted lines.
}\label{fig-ori}
\end{figure}

\begin{figure}
\includegraphics [width=145mm] {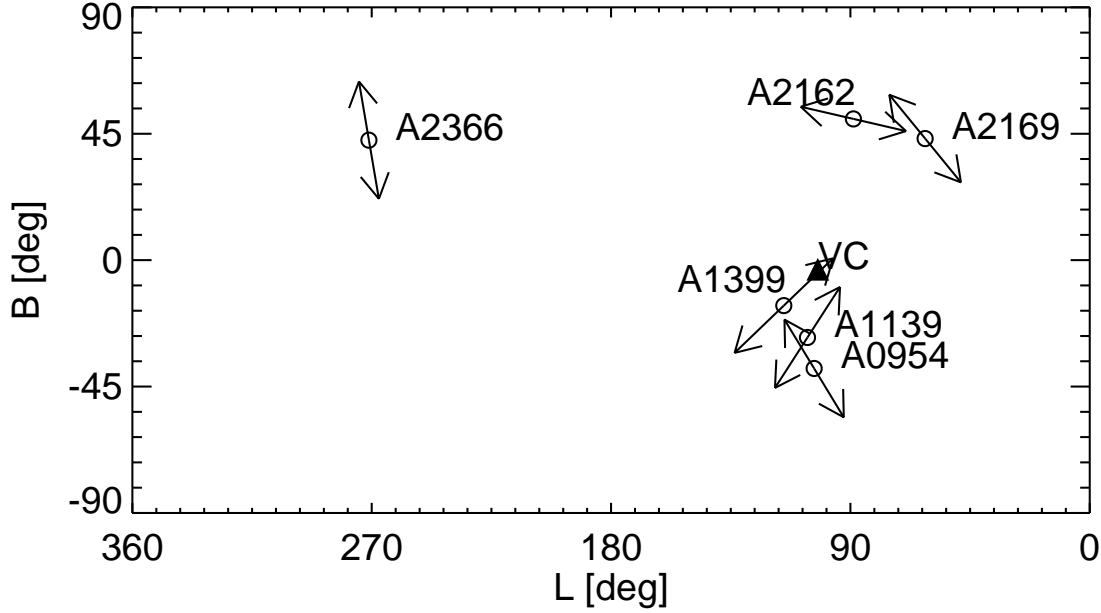}
\caption{All-sky plot of six probable rotating clusters with their rotation axes (Y1).
L and B represent supergalactic longitude and latitude, respectively. The arrow represents
the rotation axis of each cluster, and its length denotes the relative rotation amplitude, $|v_{rot}|$/${\sigma}_{p}$.
Virgo cluster is shown by a solid triangle. The LSC plane goes through the X-axis.
}\label{fig-lsc}
\end{figure}

An interesting connection between the global rotation of clusters 
and the anisotropy in galaxy alignments for clusters is expected in the context of angular momentum in galaxy clusters: 

\begin{enumerate}
\item \citet{as04,as05,as06} found that the anisotropy in galaxy alignments increases systematically from early-type (BM type I)
to late-type (BM type II-III and III) clusters. Interestingly, four of six probable rotating clusters in this study
are late-type clusters. Therefore, there may exist a connection between the rotation axes of the probable rotating clusters in this study and 
the spin-vector orientations of cluster galaxies.
In order to examine this connection, we present in Figure \ref{fig-ori} the spatial distribution and the histogram 
for the projected spin-vector orientations of galaxies
in six probable rotating clusters.
We used an orientation parameter (ORIENT) in 2dFGRS and 
an isophotal position angle parameter in $g$ band (isoPhi$\_$g) in SDSS
as the projected spin-vector orientations for cluster galaxies
\citep{col01,sto02}.
The position angles of the projected orientations for the cluster galaxies
are measured in the range of $0-180$ deg, but are plotted twice in the range of $0-360$ deg.
Figure \ref{fig-ori} shows that there is no strong anisotropy in spin-vector orientations of cluster galaxies (i.e. no strong peak in histogram).
Therefore, it is difficult to compare the position angle of the rotation axes and that of the spin-vector orientations of cluster galaxies.

\item It was found that spin-vector orientations of galaxies in clusters tend to lie parallel to the Local Supercluster (LSC) plane
\citep{as06,hu06}. In order to investigate the connection between the rotation axes of rotating clusters and the LSC plane,
we plot in Figure \ref{fig-lsc} all-sky distribution of six probable rotating clusters with their rotation axes (Y1) in the supergalactic coordinate.
It appears that there are no preferred directions of rotation axes
for six probable rotating clusters, 
while spin-vector orientations of cluster galaxies in \citet{as06} and \citet{hu06} tend to lie parallel to the Local Supercluster (LSC) plane
(the results does not change if we use Y2 instead of Y1). 
\end{enumerate}

Since we do not know the three-dimensional information about the rotation axis for rotating clusters
and the three-dimensional spin-vector orientations of cluster galaxies in this study, 
it is difficult to conclude the connection between the global rotation of clusters 
and the anisotropy in galaxy alignments for clusters at this moment.
Therefore, it is needed to study the three-dimensional spatial orientations of galaxies 
for these rotating clusters to investigate this connection.

\subsection{Cosmological implication of probable rotating clusters}
Based on the model proposed by \citet{li98} that the global rotation of the universe may provide angular momentum
to celestial bodies on their formation, \citet{god05} found the vanishing angular momenta
for the masses corresponding to the galaxy groups using Tully's galaxy groups. 
They suggested the non-vanshing angular momenta for smaller
and larger structures (compact galaxy groups and rich galaxy clusters). However, the non-vanishing angular momenta
for galaxy clusters were found indirectly based on the study of the galaxy orientation. In this study, 
we found the non-vanishing angular momenta for galaxy clusters directly based on the dynamical analysis.
This result is consistent with that of \citet{god05}. However, we found that the estimated masses for rotating clusters
are in the range of $10^{14}-10^{15} M_\odot$ (see Table \ref{tab-corr}). In addition, we could not identify
rotating clusters in massive clusters with $>2\times 10^{15} M_\odot$.
Intriguingly, the mass range of rotating clusters in this study is corresponding to the mass scale ($10^{14}-10^{15} M_\odot$)
with {\it vanishing} angular momentum predicted by \citet{god03}. Since the masses of
Tully's galaxy groups used in \citet{god05} are expected to be smaller than those of Abell clusters used in this study,
this contradiction of the mass scale between the theory and the observation should be investigated further.

\section{Summary}

We present a result of searching for galaxy clusters
that show an indication of global rotation using a spectroscopic sample of galaxies
in SDSS and 2dFGRS. Our results are summarized as follows:

\begin{enumerate}

\item We have determined the member galaxies of 899 Abell clusters covered in SDSS and 2dFGRS.
We have estimated the ratio of rotation amplitude to velocity dispersion,
and the velocity gradient for 56 clusters in which the number of member galaxies are greater than or equal to 40.

\item Among 56 clusters, we have selected 12 tentative rotating clusters
which have large values of the ratio ($>$0.53) of rotation amplitude to velocity dispersion,
and of the velocity gradient ($>$380 km s$^{-1}$ Mpc$^{-1}$).

\item We have determined the cluster morphology for 12 tentative rotating clustersmight 
using the dispersion ellipse method. The ellipticity of the dispersion ellipse 
is in the range of 0.08$-$0.57, and the position angle of major or minor axis appears
not to be related to the position angles of rotation axes (Y1 and Y2).

\item We have investigated the substructures (1D, 2D, and 3D) for the sample of tentative rotating clusters. 
DS plots show that the majority (9 out of 12) of
clusters show an evidence of substructure due to the spatially correlated
velocities of galaxies.

\item We have selected six probable rotating clusters 
(Abell 0954, Abell 1139, Abell 1399, Abell 2162, Abell 2169, and Abell 2366)
that show a single number density peak around
the cluster center with a spatial segregation of the high and low velocity galaxies
in clusters. 

\item The probable rotating clusters do not deviate significantly from the relation of 
the X-ray luminosity and the velocity dispersion or the virial mass of the clusters.
The peculiar velocities and the clustercentric distances of BCGs for two (A0954 and A1399) probable rotating clusters 
indicate that they may be in a dynamical equilibrium and undergo no recent merging.

\end{enumerate}

It will be interesting to investigate the dynamical state of probable rotating
clusters using X-ray data in order to compare the properties of the intracluster medium
and those of galaxies.

\acknowledgments
We would like to thank the referee, B. Aryal, for useful comments.
We also would like to thank all the
people involved in creating the SDSS, 2dFGRS and NED.
Funding for the SDSS and SDSS-II has been provided by the Alfred P. Sloan Foundation, 
the Participating Institutions, the National Science Foundation, 
the U.S. Department of Energy, the National Aeronautics and Space Administration, 
the Japanese Monbukagakusho, the Max Planck Society, and the Higher Education Funding Council for England.
The SDSS Web Site is http://www.sdss.org/.
The SDSS is managed by the Astrophysical Research Consortium for the Participating Institutions. 
The Participating Institutions are the American Museum of Natural History, 
Astrophysical Institute Potsdam, University of Basel, University of Cambridge, 
Case Western Reserve University, University of Chicago, Drexel University, Fermilab, 
the Institute for Advanced Study, the Japan Participation Group, Johns Hopkins University, 
the Joint Institute for Nuclear Astrophysics, the Kavli Institute for Particle Astrophysics and Cosmology, 
the Korean Scientist Group, the Chinese Academy of Sciences (LAMOST), Los Alamos National Laboratory, 
the Max-Planck-Institute for Astronomy (MPIA), the Max-Planck-Institute for Astrophysics (MPA), 
New Mexico State University, Ohio State University, University of Pittsburgh, University of Portsmouth, 
Princeton University, the United States Naval Observatory, and the University of Washington.
This research has made use of the NASA/IPAC Extragalactic Database (NED) which is 
operated by the Jet Propulsion Laboratory, California Institute of Technology, 
under contract with the National Aeronautics and Space Administration.
This work is in part supported by the grant, (R01-2004-000-10490-0) from the
Basic Research Program of the Korea Science and Engineering Foundation.

\clearpage


\clearpage

\clearpage

\clearpage

\begin{figure}
\includegraphics [width=145mm] {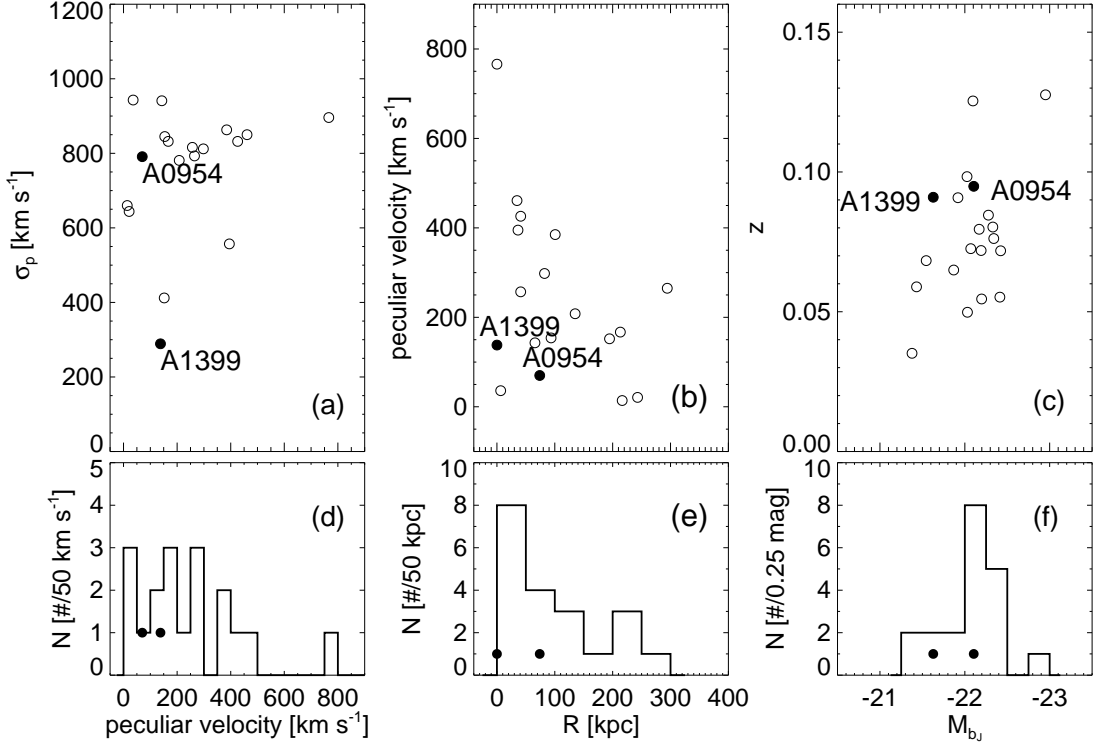}
\caption{Velocity dispersions as a function of the absolute value of
the peculiar velocity of the brightest cluster galaxies (a), redshifts as a function of the absolute magnitude of the brightest
cluster galaxies in the $b_{\rm J}$ band (b), and absolute values of peculiar velocities as a function of the clustercentric distance of
the brightest cluster galaxies (c) for two probable rotating clusters (filled circles) in comparison
with 20 non-rotating galaxy clusters (open circles) of the selected 56 galaxy clusters.
The histograms of the peculiar velocities (d), the clustercentric distances (e), and
the absolute magnitudes in the $b_{\rm J}$ band (f) of the brightest cluster galaxies are shown for 22 galaxy clusters
including two probable rotating clusters.
The two probable rotating clusters are also marked by filled circles in (d), (e), and (f).
}\label{fig-pec}
\end{figure}

\begin{table}
\scriptsize
\centering
\caption{The sample of tentative rotating clusters\label{tab-cand}}
\begin{tabular}{ccccccrcc}
\hline\hline 
 Cluster & RA     & Dec     & BM   & $\overline{cz}$ & $\sigma_p$    & $N_{gal}$ & $L_{\rm x}$ (0.1$-$2.4 keV)\tablenotemark{a}             & X-ray \\
         & [h:m:s]& [d:m:s] & Type & [km s$^{-1}$]   & [km s$^{-1}$] &     & [$10^{44}$ erg cm$^{-2}$ s$^{-1}$] & reference \\

\hline  

 S1171 & 00:01:21.70 & $-$27:32:18.0 & II      & $~ 8377_{- 117}^{+ 178}$ & $~  646_{- 166}^{+ 168}$ &   42 & ... & ...\\
 S0001 & 00:02:33.93 & $-$30:44:06.2 & I       & $~ 8815_{- 128}^{+ 142}$ & $~  577_{-  99}^{+  71}$~&   51 & ... & ...\\
 A0954 & 10:13:44.81 & $-$00:06:31.0 & ...     & $ 28459_{-  98}^{+ 108}$ & $~  801_{-  59}^{+  64}$~&   67 & 0.70 & 3\\
 A1035 & 10:32:14.16 &  ~~40:14:49.2 & II-III  & $ 21753_{- 208}^{+ 188}$ & $  1825_{-  81}^{+  82}$~&   97 & 0.92 & 2,4\\
 A1139 & 10:58:10.39 &  ~~01:35:11.0 & III     & $ 11849_{-  45}^{+  48}~$& $~  491_{-  36}^{+  37}$~&  122 & 0.09 & 1,2,3,4\\
 A1373 & 11:45:30.95 & $-$02:27:12.9 & III     & $ 37595_{- 261}^{+ 317}$ & $  1768_{- 124}^{+ 120}$ &   48 & ... & ...\\
 A1399 & 11:51:10.78 & $-$03:01:41.3 & III     & $ 27267_{-  46}^{+  48}~$& $~  289_{-  39}^{+  41}$~&   41 & ... & ...\\
 A1474 & 12:07:57.20 &  ~~14:57:18.0 & III     & $ 24151_{- 101}^{+ 105}$ & $~  714_{-  45}^{+  49}$~&   60 & 0.03 & 4\\
 A2162 & 16:12:30.00 &  ~~29:32:23.0 & II-III  & $~ 9653_{-  64}^{+  68}~$& $~  416_{-  32}^{+  33}$~&   41 & 0.01 & 4\\
 A2169 & 16:14:09.60 &  ~~49:09:10.8 & III     & $ 17343_{-  58}^{+  63}~$& $~  499_{-  39}^{+  39}$~&   71 & 0.45 & 2,4\\
 A2366 & 21:42:50.41 & $-$06:52:15.0 & I-II    & $ 15914_{-  95}^{+ 103}$ & $~  604_{-  53}^{+  56}$~&   41 & 0.05 & 4\\
 A4053 & 23:54:45.39 & $-$27:40:52.8 & III     & $ 20691_{- 184}^{+ 203}$ & $  1366_{- 143}^{+ 149}$ &   76 & ... & ...\\

\hline
\end{tabular}
\begin{flushleft}
REFERENCES: (1) \citet{bohringer00}. (2) \citet{ebeling00}. (3) \citet{bohringer04}.  (4) \citet{ledlow03}.\\
$^{\mathrm a}$ X-ray luminosity from \citet{ledlow03} is in the 0.5$-$2.0 keV band.

\end{flushleft}
\normalsize
\end{table}

\begin{table}
\centering
\caption{Kinematic properties of tentative rotating clusters\label{tab-rot}}
\begin{tabular}{cccccc}
\hline\hline 

Cluster & $\Theta_0$ & $v_{\rm rot}$     & $|v_{\rm rot}|$/${\sigma}_{p}$ & $\Theta_1$ & $dv/dR$ \\
        & (deg)      & (km s$^{-1}$) &                            & (deg)      & (km s$^{-1}$ Mpc$^{-1}$) \\
\hline  

 S1171 & $~ 11_{- 24}^{+ 22}$ & $~ 408_{- 124}^{+ 135}$ & $ 0.63_{- 0.14}^{+ 0.17}$ & $~~ 1_{- 23}^{+ 19}$ & $ 1066_{- 206}^{+ 229}$ \\
 S0001 & $ 173_{- 15}^{+ 13}$ & $~ 394_{-  78}^{+  83}~$ & $ 0.68_{- 0.10}^{+ 0.12}$ & $ 144_{- 19}^{+ 25}$ & $~ 468_{- 120}^{+ 138}$ \\
 A0954 & $ 103_{- 17}^{+ 13}$ & $~ 446_{- 133}^{+ 145}$ & $ 0.56_{- 0.16}^{+ 0.18}$ & $ 100_{- 27}^{+ 14}$ & $~ 402_{- 162}^{+ 185}$ \\
 A1035 & $ 268_{-  9}^{+ 11}$ & $ 1345_{- 177}^{+ 184}$ & $ 0.74_{- 0.09}^{+ 0.09}$ & $ 264_{-  9}^{+ 10}$ & $ 2545_{- 394}^{+ 414}$ \\
 A1139 & $~ 37_{- 11}^{+ 13}$ & $~ 288_{-  48}^{+  52}~$ & $ 0.59_{- 0.10}^{+ 0.10}$ & $~ 40_{-  9}^{+  9}~$ & $~ 464_{-  63}^{+  65}~$ \\
 A1373 & $ 155_{- 12}^{+ 11}$ & $ 1486_{- 262}^{+ 290}$ & $ 0.84_{- 0.13}^{+ 0.15}$ & $ 149_{- 17}^{+ 15}$ & $ 1923_{- 345}^{+ 448}$ \\
 A1399 & $ 207_{- 19}^{+ 27}$ & $~ 195_{-  58}^{+  62}~$ & $ 0.67_{- 0.14}^{+ 0.15}$ & $ 186_{- 14}^{+ 29}$ & $~ 418_{- 108}^{+ 151}$ \\
 A1474 & $ 268_{- 11}^{+ 15}$ & $~ 472_{- 120}^{+ 129}$ & $ 0.66_{- 0.16}^{+ 0.16}$ & $ 262_{-  8}^{+  8}~$ & $~ 789_{- 175}^{+ 193}$ \\
 A2162 & $ 359_{- 20}^{+ 20}$ & $~ 227_{-  76}^{+  80}~$ & $ 0.55_{- 0.17}^{+ 0.18}$ & $ 351_{- 32}^{+ 25}$ & $~ 426_{- 155}^{+ 177}$ \\
 A2169 & $ 342_{- 22}^{+ 19}$ & $~ 277_{-  61}^{+  61}~$ & $ 0.56_{- 0.11}^{+ 0.11}$ & $~ 23_{- 16}^{+ 13}$ & $~ 403_{- 101}^{+ 124}$ \\
 A2366 & $ 309_{- 17}^{+ 19}$ & $~ 346_{- 142}^{+ 147}$ & $ 0.57_{- 0.22}^{+ 0.21}$ & $ 300_{- 30}^{+ 19}$ & $~ 788_{- 327}^{+ 376}$ \\
 A4053 & $ 282_{- 16}^{+ 14}$ & $~ 829_{- 181}^{+ 204}$ & $ 0.61_{- 0.12}^{+ 0.12}$ & $ 269_{- 12}^{+ 15}$ & $ 1079_{- 185}^{+ 217}$ \\

\hline
\end{tabular}
\end{table}

\begin{table}
\centering
\caption{Morphological parameters of tentative rotating clusters\label{tab-mor}}
\begin{tabular}{ccccc}
\hline\hline 

Cluster & $\Gamma_A$ & $\Gamma_B$ & $\Theta_2$ & $\epsilon$ \\
        & (kpc)      & (kpc)   & (deg)      &            \\
\hline  

 S1171 & $~   537_{-     43}^{+     47}$ & $    482_{-     41}^{+     38}$ & $  159_{-  47}^{+  25}$ & $ 0.10_{- 0.09}^{+ 0.09}$ \\
 S0001 & $~   978_{-     74}^{+     72}$ & $    465_{-     62}^{+     66}$ & $~  90_{-   5}^{+   5}~$ & $ 0.52_{- 0.07}^{+ 0.07}$ \\
 A0954 & $~   993_{-     71}^{+     74}$ & $    702_{-     67}^{+     65}$ & $  133_{-  11}^{+   9}$ & $ 0.29_{- 0.08}^{+ 0.08}$ \\
 A1035 & $~   560_{-     32}^{+     34}$ & $    495_{-     34}^{+     34}$ & $  100_{-  20}^{+  28}$ & $ 0.12_{- 0.07}^{+ 0.08}$ \\
 A1139 & $~   709_{-     36}^{+     38}$ & $    620_{-     38}^{+     37}$ & $~  74_{-  16}^{+  19}$ & $ 0.13_{- 0.07}^{+ 0.07}$ \\
 A1373 & $~   689_{-     56}^{+     61}$ & $    545_{-     45}^{+     48}$ & $  170_{-  15}^{+  20}$ & $ 0.21_{- 0.10}^{+ 0.09}$ \\
 A1399 & $~   772_{-     74}^{+     74}$ & $    334_{-     32}^{+     32}$ & $  138_{-   5}^{+   5}~$ & $ 0.57_{- 0.06}^{+ 0.06}$ \\
 A1474 & $~   912_{-     62}^{+     61}$ & $    563_{-     43}^{+     45}$ & $~  69_{-   6}^{+   7}~$ & $ 0.38_{- 0.07}^{+ 0.07}$ \\
 A2162 & $~   502_{-     47}^{+     51}$ & $    461_{-     50}^{+     48}$ & $~  44_{-  42}^{+  49}$ & $ 0.08_{- 0.09}^{+ 0.11}$ \\
 A2169 & $~   941_{-     57}^{+     56}$ & $    631_{-     56}^{+     57}$ & $~  58_{-   8}^{+   9}~$ & $ 0.33_{- 0.07}^{+ 0.07}$ \\
 A2366 & $~   419_{-     39}^{+     42}$ & $    324_{-     36}^{+     37}$ & $  148_{-  22}^{+  18}$ & $ 0.23_{- 0.10}^{+ 0.09}$ \\
 A4053 & $   1061_{-     73}^{+     70}$ & $    676_{-     53}^{+     56}$ & $~  51_{-   9}^{+   6}~$ & $ 0.36_{- 0.07}^{+ 0.07}$ \\

\hline
\end{tabular}
\end{table}

\begin{table}
\scriptsize
\centering
\caption{Parameters for testing the presence of substructure\label{tab-sub}}
\begin{tabular}{rrrrrrrrrrrrc}
\hline\hline 
Cluster & $I$ & $I_{90}$ & Skewness & clRej & Kurtosis & clRej & AI & clRej & TI & clRej & $\Delta_{obs}$ & f($\Delta_{sim}>\Delta_{obs}$) \\
\hline  

 S1171 &  1.21 &  1.11 & $  0.96$ &   99.3 & $ -0.45 $ &  10.0 & $  0.58 $ &  73.0 &   1.57 &   98.3 &    68 &  0.335 \\
 S0001 &  0.93 &  1.09 & $ -0.45$ &   86.3 & $ -0.92 $ &  89.9 & $ -0.08 $ &  13.0 &   0.91 &   53.8 &    98 &  0.000 \\
 A0954 &  0.97 &  1.07 & $ -0.29$ &   71.8 & $ -0.64 $ &  68.4 & $ -0.44 $ &  61.1 &   0.92 &   52.8 &    97 &  0.006 \\
 A1035 &  1.00 &  1.05 & $  0.34$ &   85.3 & $ -1.24 $ & 100.0 & $  1.07 $ &  96.4 &   0.74 &   99.9 &   167 &  0.000 \\
 A1139 &  1.02 &  1.04 & $ -0.18$ &   60.7 & $  0.23 $ &  64.1 & $  0.29 $ &  42.7 &   0.94 &   47.8 &   206 &  0.000 \\
 A1373 &  0.99 &  1.10 & $ -0.27$ &   61.9 & $ -1.19 $ &  99.3 & $ -0.82 $ &  88.9 &   0.82 &   87.8 &    90 &  0.000 \\
 A1399 &  1.07 &  1.11 & $ -0.50$ &   86.3 & $  0.29 $ &  70.8 & $ -0.62 $ &  76.2 &   0.90 &   52.5 &    75 &  0.001 \\
 A1474 &  0.95 &  1.08 & $  0.41$ &   83.3 & $ -1.00 $ &  97.0 & $  1.36 $ &  99.2 &   0.73 &   99.5 &    84 &  0.012 \\
 A2162 &  0.92 &  1.11 & $  0.03$ &    6.9 & $ -1.11 $ &  95.9 & $ -0.34 $ &  48.9 &   0.86 &   71.7 &    42 &  0.464 \\
 A2169 &  1.01 &  1.07 & $ -0.45$ &   91.1 & $  0.07 $ &  48.8 & $ -0.78 $ &  86.7 &   0.88 &   72.5 &   120 &  0.000 \\
 A2366 &  0.95 &  1.11 & $  0.26$ &   57.4 & $ -0.73 $ &  56.8 & $  0.63 $ &  77.0 &   0.78 &   93.9 &    41 &  0.707 \\
 A4053 &  1.20 &  1.07 & $ -1.02$ &   99.9 & $  0.57 $ &  84.3 & $ -1.59 $ &  99.8 &   0.87 &   78.6 &   174 &  0.000 \\

\hline
\end{tabular}
\normalsize
\end{table}

\begin{table}
\centering
\caption{Summary of global kinematic properties for tentative rotating clusters\label{tab-sum}}
\begin{tabular}{ccccccc}
\hline\hline 
Cluster & \multicolumn{3}{c}{Substructure} & Morphology & Dynamical   & Probable rotating \\
\cline{2-4}
        & 1D & 2D & 3D                     &            & equilibrium & cluster \\
\hline  

 S1171 &  Yes & Yes & No                   & Spherical  & ...         & No     \\
 S0001 &  No  & Yes & Yes                  & Elongated? & ...         & No      \\
 A0954 &  No  & Yes & Yes                  & Elongated  & Yes         & Yes     \\
 A1035 &  Yes & Yes & Yes                  & Spherical  & ...         & No      \\
 A1139 &  No  & No  & Yes                  & Spherical  & Yes         & Yes     \\
 A1373 &  Yes & Yes & Yes                  & Elongated  & ...         & No      \\
 A1399 &  No  & No  & Yes                  & Elongated  & Yes         & Yes     \\
 A1474 &  Yes & Yes & Yes                  & Elongated  & ...         & No      \\
 A2162 &  Yes & No  & No                   & Spherical  & Yes         & Yes     \\
 A2169 &  No  & No  & Yes                  & Elongated  & Yes         & Yes     \\
 A2366 &  No  & No  & No                   & Elongated  & Yes         & Yes     \\
 A4053 &  Yes & No  & Yes                  & Elongated  & ...         & No      \\

\hline
\end{tabular}
\normalsize
\end{table}

\begin{table}
\centering
\caption{Velocity dispersions and virial masses for probable rotating clusters\label{tab-corr}}
\begin{tabular}{ccccc}
\hline\hline 

Cluster & ${\sigma}_{p}$ & ${\sigma}_{p,r}$ & M$_{vir}$         & M$_{vir,r}$ \\
        &  (km s$^{-1}$) &    (km s$^{-1}$) & [$10^{14} M_\odot$] & [$10^{14} M_\odot$] \\ 
\hline  

 A0954 & $   801_{-  62}^{+  63}$ & $   719_{-  67}^{+  74}$ & $ 11.65_{-  1.73}^{+  1.92}$ & $  9.39_{-  1.67}^{+  2.03}$ \\
 A1139 & $   491_{-  32}^{+  39}$ & $   423_{-  36}^{+  40}$ & $~ 3.93_{-  0.49}^{+  0.66}$ & $  2.92_{-  0.48}^{+  0.58}$ \\
 A1399 & $   289_{-  39}^{+  43}$ & $   248_{-  38}^{+  44}$ & $~ 1.06_{-  0.27}^{+  0.34}$ & $  0.77_{-  0.22}^{+  0.30}$ \\
 A2162 & $   416_{-  31}^{+  32}$ & $   377_{-  39}^{+  40}$ & $~ 1.67_{-  0.24}^{+  0.26}$ & $  1.37_{-  0.27}^{+  0.31}$ \\
 A2169 & $   499_{-  40}^{+  42}$ & $   431_{-  43}^{+  45}$ & $~ 4.78_{-  0.74}^{+  0.83}$ & $  3.56_{-  0.67}^{+  0.78}$ \\
 A2366 & $   604_{-  53}^{+  58}$ & $   569_{-  62}^{+  65}$ & $~ 3.45_{-  0.58}^{+  0.69}$ & $  3.06_{-  0.63}^{+  0.73}$ \\

\hline
\end{tabular}
\normalsize
\end{table}


\end{document}